# Transitive inference as probabilistic preference learning

Francesco Mannella, Giovanni Pezzulo

*Institute of Cognitive Sciences and Technologies, National Research Council, Rome, Italy*

**Corresponding author**

Giovanni Pezzulo

Institute of Cognitive Sciences and Technologies,

National Research Council,

00185 Rome, Italy

E-mail: giovanni.pezzulo@istc.cnr.it

**Abstract**

Transitive Inference (TI) is a cognitive task that assesses an organism's ability to infer novel relations between items based on previously acquired knowledge. TI is known for exhibiting various behavioral and neural signatures, such as the Serial Position Effect (SPE), Symbolic Distance Effect (SDE), and the brain's capacity to maintain and merge separate ranking models. We propose a novel framework that casts TI as a probabilistic preference learning task, using one-parameter Mallows models. We present a series of simulations that highlight the effectiveness of our novel approach. We show that the Mallows ranking model natively reproduces SDE and SPE. Furthermore, extending the model using Bayesian selection showcases its capacity to generate and merge ranking hypotheses as pairs with connecting symbols are encountered. Finally, we employ neural networks to replicate Mallows models, demonstrating how this framework aligns with observed prefrontal neural activity during TI. Our innovative approach sheds new light on the nature of TI, emphasizing the potential of probabilistic preference learning for unraveling its underlying neural mechanisms.

**Introduction**

When we are informed that Alice is taller than Brenda, and Brenda is taller than Cindy, we possess the capacity to swiftly deduce that Alice is indeed taller than Cindy, even in the absence of explicit information to that effect. This cognitive ability, referred to as Transitive Inference (TI), is believed to be present in both humans and other animals. It is thought to rely on the establishment of relational networks or mental "schemas" that facilitate the coherent organization of experiences and enable the inference of previously unencountered events (Bryant and Trabasso, 1971; Jensen et al., 2019; McGonigle and Chalmers, 1977; Piaget, 1947).

During a typical laboratory test of TI, participants firstly learn pairwise comparisons between items (e.g., A > B, B > C, and C > D) and then are asked to perform novel pairwise comparisons (e.g., A > C and A > D) that require using what they previously learned. This is possible to the extent that participants map all the learned items (e.g., A, B, C, D) in a ranked mental schema (e.g., a representation that A > B > C > D etc.) that affords pairwise comparisons between seen or unseen pairs, by simply considering which symbol comes first in the ranked mental schema.

This experimental paradigm has been widely utilized to investigate Transitive Inference and explore its neural foundations in various species, including humans (Bryant and Trabasso, 1971; Burt, 1909; Merritt and Terrace, 2011; Zeithamova et al., 2012), primates (Brunamonti et al., 2016; Treichler et al., 2003; Treichler and Van Tilburg, 1996), rodents (Davis, 1992; Dusek and Eichenbaum, 1997), and birds (Bond et al., 2010, 2003; Lazareva et al., 2015; Lazareva and Wasserman, 2012; Wei et al., 2014).

Figure 1 presents a visual summary of the key findings derived from numerous studies in this field, highlighting several notable phenomena. At the behavioral level, researchers typically observe two fundamental effects across a wide array of investigations: the Symbolic Distance Effect (SDE) and the Serial Position Effect (SPE). The SDE is characterized by the observation that performance improves linearly with an increase in the symbolic distance between the items being compared. In practical terms, this means that comparing, for instance, B with D is inherently easier than comparing B with C, given the greater symbolic distance between B and D. Conversely, the SPE, also known as the "terminal item" effect, indicates that performance is enhanced when the comparison involves symbols located at the ends of a sequence. For instance, in a list of symbols comprising A, B, C, D, and E, comparisons that include terminal symbols, like AB and CE, tend to be more efficiently executed than comparisons involving symbols situated in the middle, such as BC. This phenomenon underscores the influence of the position of items within the sequence on cognitive processing.

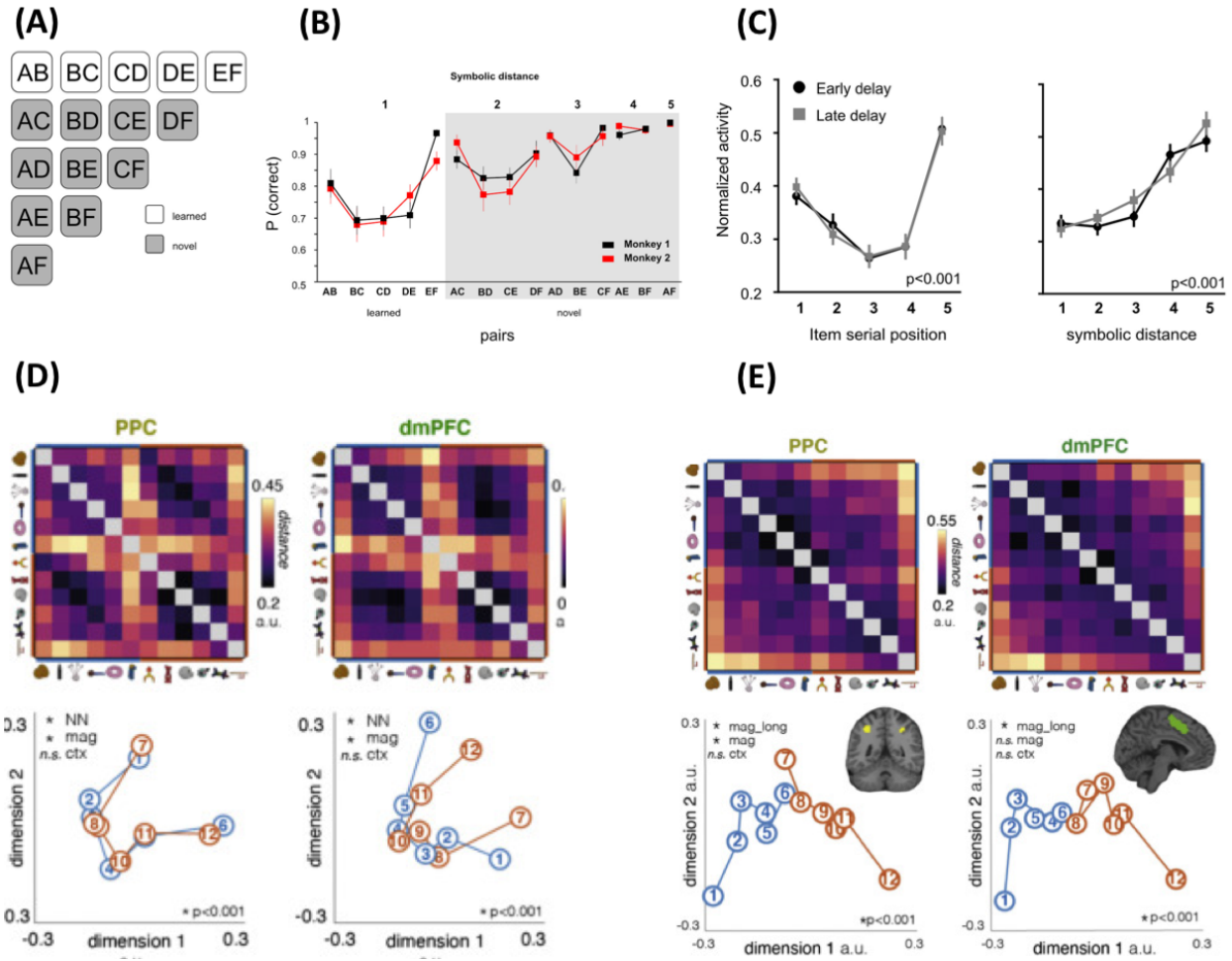

***Figure 1. Graphical Representation of a Typical Transitive Inference (TI) Task and Key Empirical Observations, Spanning Behavioral (A-B) and Neural (C-E) Levels.*** *(A) In a conventional TI paradigm, participants are presented with pairs in white and provided with instructions for the correct responses (e.g., A > B, B > C). Subsequently, their performance is evaluated through the presentation of novel pairs in gray. (B) Response patterns of monkeys engaged in a TI task, adapted from (Brunamonti et al., 2016). This graph illustrates the increase in the probability of a correct response as the symbolic distance between symbols in a pair widens (Symbolic Distance Effect, SDE), and the influence of the pair including one of the terminal symbols A or F (Serial Position Effect, SPE). (C) Neural substrates associated with TI in the prefrontal cortex of monkeys, based on data from (Brunamonti et al., 2016). The figure displays both neural and behavioral modulation pertaining to SPE (left) and SDE (right). It includes the mean and Standard Error of the Mean (SEM) for the performance of Monkey 1 (top, black lines) and Monkey 2 (top, red lines), along with normalized population neural activity in the early (bottom, black) and late (bottom, gray) delays for each serial position (left) and symbolic distance (right). (D-E) Neural observations from (Nelli et al., 2023) reveal that following "boundary training trials," neural representations in the posterior parietal cortex (PPC) and dorsomedial prefrontal cortex (dmPFC) rapidly link two previously separate rankings into a single, coherent ranking. See the main text for further information.*

Intriguingly, at the neural level, a study involving monkeys reported that neurons recorded in the prefrontal cortex exhibit tuning for both the Symbolic Distance Effect (SDE) and the Serial Position Effect (SPE) (Brunamonti et al., 2016). Note that this tuning is not a group-level effect, since it is present also in single monkeys. Furthermore, recent functional magnetic resonance imaging (fMRI) research has provided valuable insights into the development of neural codes underlying Transitive Inference (TI). This study revealed that, during the acquisition of a novel TI task, neural representations in frontoparietal brain regions, encoding two initially disconnected rankings (e.g., ABC and DEF), swiftly reorganize into a single ranking following brief exposure to linking information (e.g., C>D) (Nelli et al., 2023). This investigation underscores the idea that mastering a TI task entails not only deducing the relative positions of items within a ranking but also understanding the number of rankings and their respective lengths. Moreover, the results of this study (and others (Xie et al., 2022; Xu et al., 2024)) demonstrate the rapid adaptability of frontoparietal neural populations in supporting these learning processes.

From a computational perspective, explaining Transitive Inference (TI) has been addressed using various reinforcement-based models, which use mechanisms such as associative learning, value transfer and reinforcement (Ciranka et al., 2022; Lazareva and Wasserman, 2012); see (Vasconcelos, 2008) for a comprehensive review. More recently, various cognitive models have emerged, which explicitly represent the order of a set of items, as emphasized by (Jensen et al., 2013). One notable example of such a cognitive model is the Betasort model. This probabilistic model represents the positions of stimuli along a spatial line using beta distributions and dynamically updates these positions based on positive and negative feedback from correct and incorrect responses in each trial (Jensen et al., 2015). Another effective approach to TI involves the use of neural networks (Botvinick and Watanabe, 2007; Siemann and Delius, 1998). For example, an early study demonstrated that that a simple three-layer network can simulate symbolic distance effect (De Lillo et al., 2001). More recent studies demonstrated that after training on seen pairs (e.g., AB, BC, CD) with gradient descent optimization and backpropagation through time, most types of recurrent neural networks have demonstrated success in generalizing to unseen pairs (e.g., AC, BD) (Di Antonio et al., 2023; Kay et al., 2024).

These computational models have shown the ability to replicate the observed Symbolic Distance Effect and/or Serial Position Effect. However, they have yet to be tested on tasks that require the integration of previously separated rankings into a single ranking, as reported by (Nelli et al., 2023). The discovery that a brief exposure to linking information can effectively merge disjoint rankings into a unified one poses a challenge for models like Betasort, which assume the presence of a single ranking from the outset. In contrast, a two-layer feedforward neural network model, trained using stochastic gradient descent, has been able to replicate this phenomenon (Nelli et al., 2023), but it was not directly used to reproduce the Serial Position Effect. Then, despite the achievements of these previous approaches, we still lack a comprehensive computational framework that can simultaneously account for the full spectrum of findings in the TI literature in a parsimonious way.

Here, we present a novel and integrative theory for Transitive Inference (TI) termed "probabilistic preference learning." This theory offers a coherent explanation of the empirical findings discussed above. The theory posits that the brain accomplishes TI by acquiring a parametric generative model of rankings, often referred to as a "mental schema." This schema encodes a probability distribution of rankings for the symbols observed, meaning that, after witnessing relationships like A>B, B>C, C>D, and D>E, the model calculates the probabilities associated with symbol sequences like AB, BC, CD, ABCD, ADCB, ACBD, ABCDE, ACDEB, and so forth. Following the learning process, the ranking model enables the brain to make TI-based decisions for both familiar pairs (e.g., A and B) and novel pairs (e.g., A and C) by simply inferring the probabilities of these pairs (e.g., P(AB) or P(AC)). This novel framework offers a unified and comprehensive account of TI, merging probabilistic modeling and preference learning to elucidate the underlying cognitive processes.

To empirically validate our theory, we have applied it using one of the earliest and more straightforward parametric generative models for rankings known as the Mallows model (Mallows, 1957). Our study comprises three distinct simulations, each serving a specific purpose. The first simulation focuses on elucidating the dynamics of inference, the second simulation addresses the dynamics of learning, and the third simulation explores a potential neural network implementation of our proposed model. Subsequently, we compare the outcomes of these simulations with the critical behavioral aspects (first and second simulations) as well as the neural signatures of TI (third simulation), illustrated in Figure 1. This rigorous comparative analysis allows us to gauge the effectiveness of our theory in replicating key behavioral and neural observations associated with Transitive Inference.

To provide a glimpse of our findings, the first simulation demonstrates that within a trained Mallows model, decisions regarding both familiar and unfamiliar pairs can be readily computed by evaluating the pair probabilities within the model (e.g., P(AB) or P(AC)). This approach effectively replicates both the Symbolic Distance Effect (SDE) and the Serial Position Effect (SPE).

In the second simulation, we reveal that when the Mallows model is incrementally trained to solve a Transitive Inference (TI) task involving initially separate rankings and is subsequently exposed to a connecting pair, it emulates the rapid process of "knowledge assembly." This phenomenon mirrors the observed integration of novel items into coherent transitive orderings, akin to the cognitive processes observed in human frontoparietal areas, as reported in (Nelli et al., 2023).

Lastly, the third simulation underscores that a neural network trained to perform TI based on the Mallows model develops internal representations that align with the prefrontal coding observed in primates, as in (Brunamonti et al., 2016) and in humans, as in (Nelli et al., 2023). This convergence of computational and neural model results further supports the robustness and validity of our proposed probabilistic preference learning theory.

**Methods**

*The Mallows ranking model*

The Mallows ranking model affords probabilistic preference learning and inference by capturing the probability distribution of potential rankings for observed symbols (Mallows, 1957). This model is characterized as an exponential distribution, which gauges the distances of all ranking permutations from a central point:

$$p(\tau_i|\beta, \mathbf{c}) \approx e^{-\beta \, \mathcal{K}(\tau_i, \mathbf{c})}$$

Here, $\tau_i \in \{\tau_1, \tau_2, \ldots, \tau_N\}$ represents a given ranking sequence within the set $\mathcal{P}$ of all ranking permutations (for instance, ACDBEF or CBDFEA); $\mathbf{c}$ stands for the central point of the distribution, which is the ranking considered to have the highest probability (e.g., ABCDEF); and $\beta$ is the sole model parameter, determining the precision of the model or the degree of uncertainty around the central point.

Figure 2A offers a visual example of the Mallows ranking model. In this instance, the underlying distribution is assumed to be Gaussian, and the central point is set as ABCDEF. As depicted in the figure, the model assigns the highest probability to the central point and progressively lower probabilities to each of the potential permutations (e.g., ACDBEF). Notably, the probability associated with a specific ranking permutation (e.g., ACDBEF) is contingent on its distance from the central point (e.g., ABCDEF).

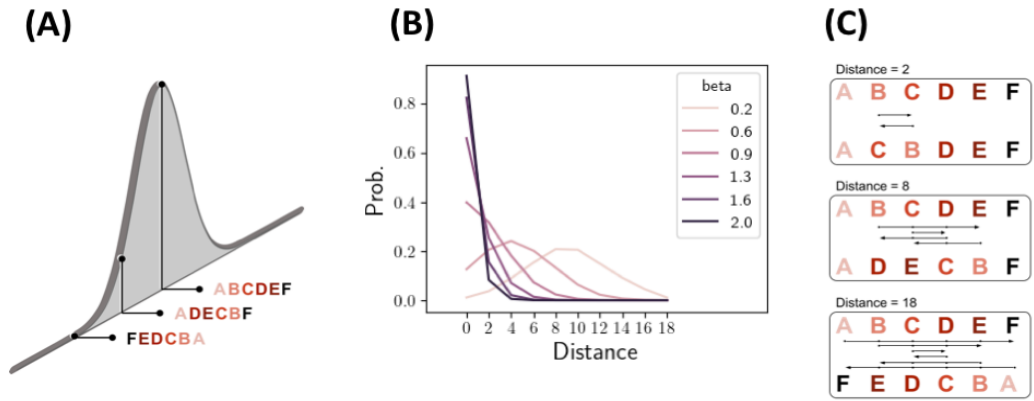

*Figure 2. Visualization of the Mallows Ranking Model. (A) A graphical representation of the Mallows ranking model, assuming a Gaussian distribution with the central point set at ABCDEF. Notably, for values of beta smaller than one, the central point exhibits the highest probability, while the probabilities assigned to other permutations decrease as they move farther away from the central point. For illustrative purposes, we also show values of beta greater than one, for which the model assigns the highest probability to rankings that differ from the central point, possibly reflecting learning errors or cognitive deficits. (B) Probabilities assigned to all permutations, varying in Spearman's footrule*

*distance from 0 to 18 concerning the central point. (C) Instances of permutations and their respective distances from the central point (ABCDEF). Further details can be found in the main text.*

The computation of this distance necessitates the selection of a metric $\mathcal{K}$. In this context, we employ one of the most commonly used metrics, namely Spearman's footrule distance (Diaconis, 1988), which is defined as follows:

$$\sum_i |\tau_i - c_i|$$

In Figure 2B, we display the probabilities assigned to all possible permutations of 6-symbol sequences, where their Spearman's footrule distance from the central point spans from 0 (when the permutation is identical to the central point, e.g., ABCDEF) to 18 (e.g., FEDCBA), considering various choices for the parameter $\beta$. Additional examples are presented in Figure 2C, and further details and results for alternative distances and parameter values can be found in the Appendix. The operation of this model will be elucidated in Simulation 1 as described below.

The Mallows model enables a relatively simple form of statistical learning, involving the continuous refinement of the estimate for the central point following the observation of each pair (Mallows, 1957). When provided with a sample of permutations $\mathbf{r}_k \in \{\mathbf{r}_1, \mathbf{r}_2, \ldots, \mathbf{r}_K\}$, the maximum likelihood estimation (MLE) for the central point can be calculated as follows:

$$\mathbf{c}_{MLE} = \arg\min_{\mathbf{c} \in \mathcal{P}} \sum_k \mathcal{K}(\mathbf{r}_i, \mathbf{c})$$

**Bayesian Model Selection for Incremental Learning**

The Mallows model is premised on the assumption that the length of the ranking coincides with the number of observed symbols. For instance, after encountering 3 symbols in 2 pairs (e.g., A > B and B > C), the model presumes the existence of a single ranking comprising 3 symbols (e.g., ABC). However, in the context of Transitive Inference (TI) tasks, the number of observed symbols is dynamic, evolving as novel pairs are introduced. Furthermore, in certain scenarios, it may be challenging to seamlessly amalgamate the observed symbols into a single ranking (e.g., when observing A > B and C > D).

In such situations, individuals experience not only uncertainty regarding the correct ranking but also regarding the presence of either a single ranking or multiple rankings and the number of symbols within each ranking. To address this uncertainty, it is proposed that individuals do not consider a single Mallows model (and its associated ranking), but rather a probability distribution encompassing various hypotheses. Each hypothesis corresponds to one or more Mallows models, differing in the number of symbols they involve.

For example, after observing the same 3 symbols in 2 pairs as mentioned earlier (e.g., A > B and B > C), an individual might entertain two hypotheses. The first hypothesis posits that the observed pairs can be explained by a single Mallows model comprising three symbols (denoted as [ABC]). The second hypothesis suggests that the observed pairs can be explained by two Mallows models, each with two symbols (referred to as [AB] and [BC]).

To choose between these hypotheses, individuals may strive to strike a balance between accuracy (i.e., the ability to account for more data) and complexity (i.e., the economy of parameters), as formally described by the Bayesian Information Criterion (BIC) (Schwarz, 1978). In the example above, both hypotheses exhibit the same level of accuracy since they elucidate all the observed data. However, the former hypothesis entails less complexity, as it requires only one model as opposed to two models, resulting in a lower BIC score. It is posited that individuals would opt for the former hypothesis, which boasts a lower BIC score.

This innovative extension of the Mallows model enables individuals to engage in incremental learning and the flexibility to reconsider their ranking model as new data becomes available. Simulation 2, outlined below, will provide a practical demonstration of this concept.

**Neural Network Implementation of the Mallows Ranking Model for In-Depth Neural-Level Analysis**

To facilitate a neural-level investigation, we've devised a straightforward neural network architecture, a multilayer perceptron (MLP), designed to learn neural-like codes that mimic the probability distributions of the Mallows model (see Figure 3A).

To construct the training data, we randomly sampled 1500 permutations from the Mallows distribution described in the first simulation (refer to Figure 4). For each of these permutations, we generated 15 target outputs, corresponding to the possible pairs such as AB, BC, … AF. For example, if the sampled permutation is [BACDFE], the correct target for the pair AB is 0 (FALSE) since A follows B in the current permutation, while the correct target for the pair BC is 1 (TRUE) as B precedes C in the current permutation, and so on. (See Figure 3B for a visual representation of the 15 targets for the permutation [BACDFE]). The MLP is trained in a supervised manner using the dataset outlined above to produce the appropriate target (e.g., A) when given each pair as input (e.g., AC). This training minimizes the mean square error between the output and the correct target. Over time, the MLP learns to generate Transitive Inference (TI) responses that are virtually indistinguishable from those generated by the Mallows model, which served as the source of the dataset. This enables us to examine the neural-like codes in the hidden layer of the MLP that produce these responses, as demonstrated in the third simulation.

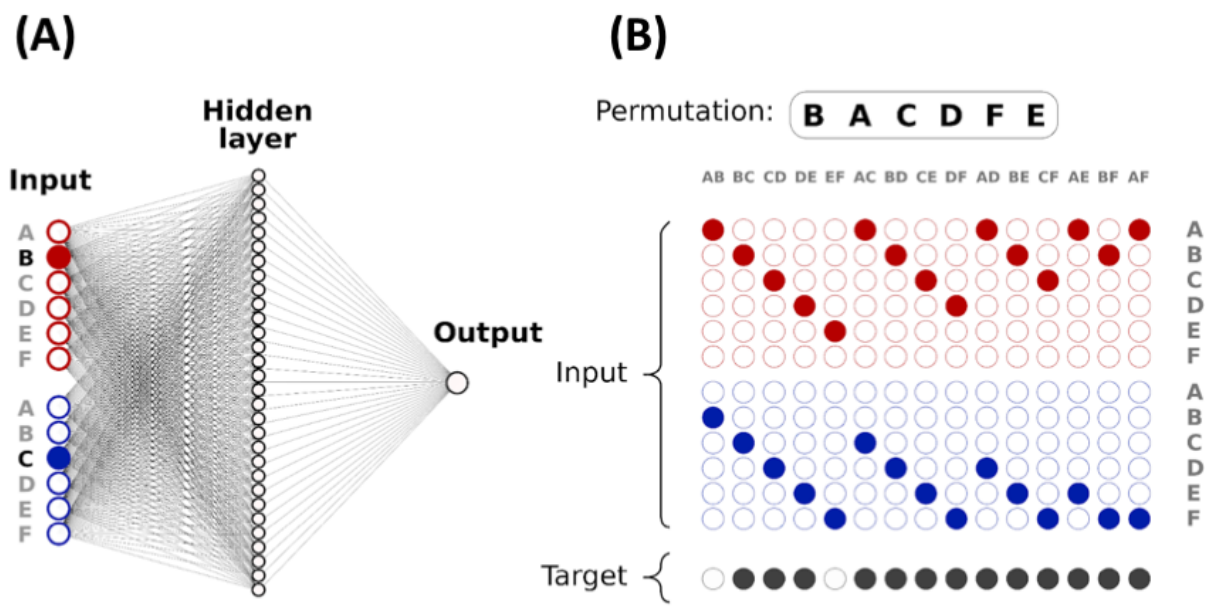

*Figure 3. Neural network implementation of the Mallows model for neural-level analysis. (A) The neural model's architecture is a multilayer perceptron (MLP) featuring 12 input nodes, 30 hidden nodes, and 1 output node. The input layer receives a concatenated representation of two symbols, with each symbol expressed as a one-hot vector consisting of 6 elements (a categorical variable where one value is 1 while others are 0). The hidden layer comprises 30 sigmoid units, and the output layer consists of a single sigmoid unit, with its activation determining the predicted probability of selecting the first symbol. (B) An example illustrating patterns with their corresponding target responses, drawn from a sample permutation [BACDFE]. In this example, the correct target for the pair AB is 0 (FALSE) as A follows B in the current permutation, while the correct target for the pair BC is 1 (TRUE) since B precedes C in the current permutation, and so forth. The training dataset is generated by randomly sampling 1500 permutations from the Mallows distribution, as depicted in Figure 4. Additional details are provided in the main text.*

**Results**

Our comprehensive investigation encompasses three simulations, each focused on distinct facets of the Mallows ranking model. The first simulation delves into the decision dynamics exhibited by a well-trained Mallows ranking model. The second simulation unravels the learning dynamics inherent in the model when confronted with a novel Transitive Inference (TI) task. Finally, the third simulation sheds light on the neural-like codes harnessed by the model to carry out TI, offering a comprehensive view of this cognitive process.

**First Simulation: Decision Dynamics of the Mallows Ranking Model**

The first simulation provides an insight into how a well-trained 6-symbol Mallows ranking model, with ABCDEF as the central point, determines which of two symbols, A or B, holds a higher ranking.

This decision can be made by calculating the probability that A outranks B (i.e., P(AB)) across all ranking permutations of the model:

$$p(AB) = \frac{\sum p(\boldsymbol{\tau}_{A>B})}{\sum p(\boldsymbol{\tau}_{A>B}) + \sum p(\boldsymbol{\tau}_{B>A})}$$

The model subsequently selects A if P(AB) exceeds 0.5, and B otherwise.

In essence, the model can estimate the probability that symbol A possesses a higher rank than symbol B by taking into account all ranking permutations (or a subset of them, as discussed below) and tallying how frequently symbol A appears in a higher rank position than symbol B. Given that ranking permutations closer to the central point carry a higher probability, they exert a more substantial influence on the decision-making process.

Figure 4A presents the outcomes of a simulation for a 6-symbol Mallows model with ABCDEF as the central point and a precision of = 0.8 (this value is chosen for illustrative purposes; please refer to the Appendix for a discussion of results with alternative distance metrics and parameter values). The figure depicts the choice probabilities for all 15 pairs (e.g., AB, AC, AD, etc.) achievable with 6 symbols. The symbolic distance between these pairs is indicated in the legend and varies from 1 (e.g., AB or BC) to 5 (i.e., AF). As observed in the figure, the simulation effectively replicates the two pivotal features of Transitive Inference (TI) identified in human and animal studies: the Symbolic Distance Effect (SDE), wherein performance improves with increasing symbolic distance between symbols, and the Serial Position Effect (SPE), indicating enhanced performance when comparisons involve the terminal symbols.

Figure 4B offers the results of the same simulation, but in this case, probabilities are computed using an approximate Bayesian approach. Specifically, this involves initially sampling 1500 permutations randomly from the Mallows model and then deriving the probability P(AB) solely from these sampled permutations. The results stemming from this approximate, sampling-based Bayesian procedure essentially mirror those of the full Bayesian procedure presented in Figure 4A, with both the SDE and SPE effects remaining apparent. The pattern of results would essentially remain the same, even with the inclusion of other sources of noise (e.g., stochasticity in the decision process), as long as the noise remains within reasonable limits.

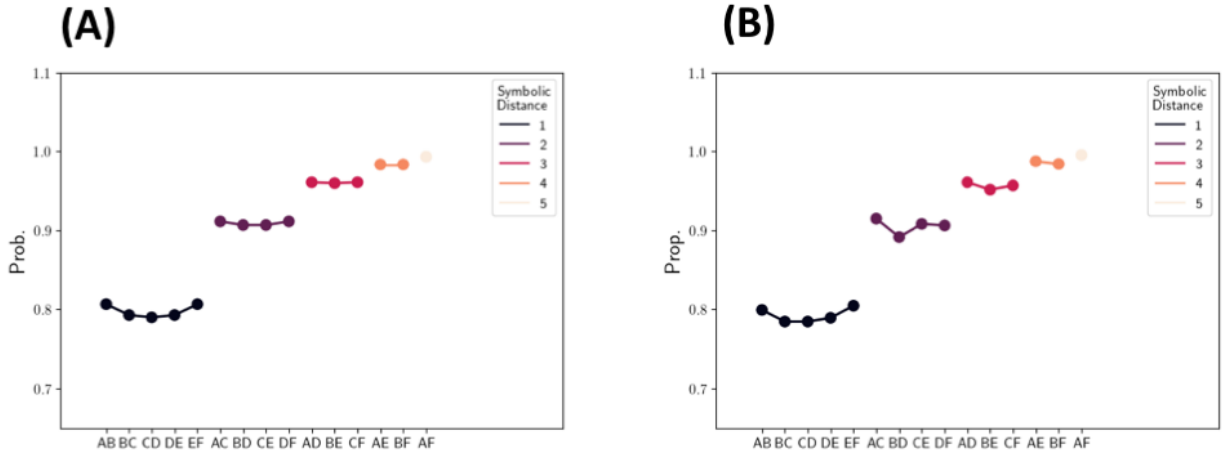

***Figure 4. Results of the first simulation.*** *(A) Probability of relative-rank decisions between all pairs of symbols (e.g., AB), with the sequence ABCDEF as the central point and $\beta$ = 0.8. It is worth noting that this plot successfully reproduces both the Symbolic Distance Effect (SDE) and the Serial Position Effect (SPE). (B) Probability of relative-rank positions between all pairs of symbols, determined by considering 1500 randomly sampled permutations. Further details are provided in the main text.*

**Second Simulation: Learning Dynamics of the Mallows Ranking Model**

The second simulation is designed to showcase how the Mallows ranking model can be trained to perform Transitive Inference (TI) using the Bayesian model selection procedure, as detailed in the Methods section. This procedure enables the model to assess which Mallows model (or combination of models) best accounts for the observed data. The training unfolds across five stages, with each stage introducing a sequence of pairs as follows: A > B, B > C, D > E, E > F, and lastly C > D. Of note is the initial isolation of the two sets of symbols, ABC and DEF, which remain disconnected until the fifth training stage when the connecting pair (C > D) is introduced. This setup allows for an examination of whether the model can connect rankings, a phenomenon reported in empirical studies (Nelli et al., 2023).

Figure 5 offers an overview of the simulation's results. The plot depicts the Bayesian Information Criterion (BIC) scores for various competing hypotheses (e.g., here, 25 hypotheses are considered) throughout the experiment. These hypotheses encompass single Mallows models (e.g., [AB]) or combinations of two or more Mallows models (e.g., [AB] - [BC], [AB] - [BC] - [DE], etc.). The results are organized into five rows, with each row corresponding to a training stage. The rows show the BIC scores assigned to the competing hypotheses after observing a single pair (A > B, first row), two pairs (A > B and B > C, second row), and so on. For ease of visualization, the scores are normalized between 0 and 1. The winning model (i.e., the model with the highest BIC score) for each row is marked with an asterisk. For simplicity, models with zero accuracy across all training stages are not displayed.

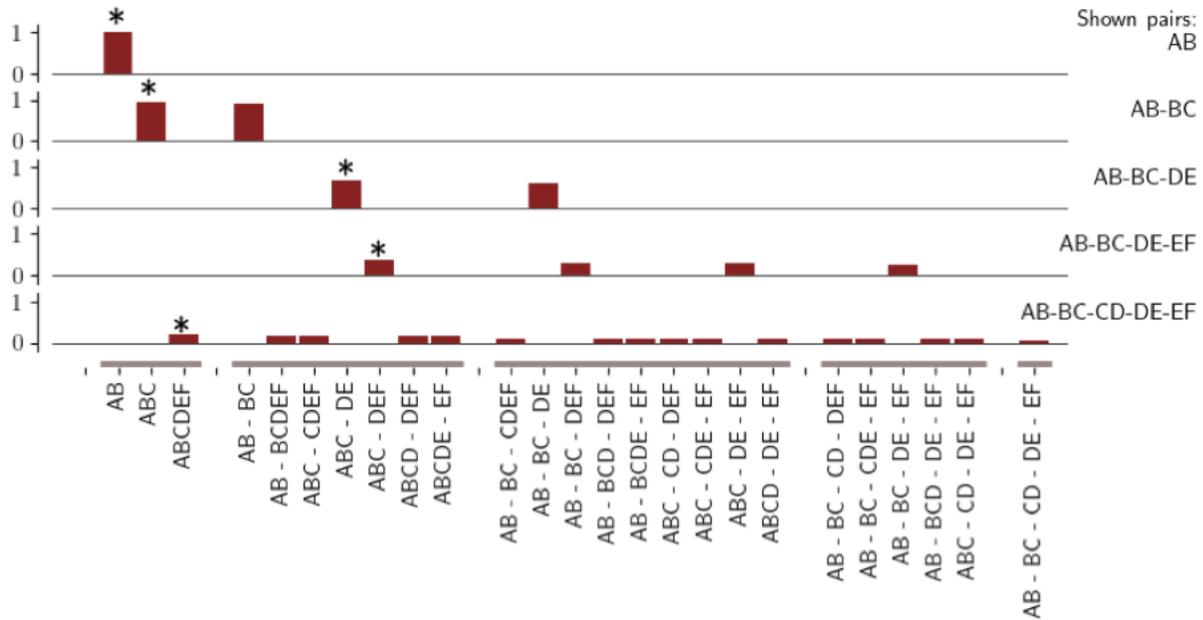

***Figure 5. Results of the Second Simulation.*** *This plot displays the (inverse) Bayesian Information Criterion (BIC) scores for various competing hypotheses, including single Mallows models like [AB] and combinations of two or more Mallows models such as [AB] - [BC], [AB] - [BC] - [DE], and so on, as the training progresses and new pairs of symbols are introduced. To enhance clarity, the BIC scores have been normalized between 0 and 1, with 1 representing the best score. The results are organized into five rows, with each row corresponding to a training stage. The winning model (i.e., the one with the highest BIC score) for each stage is indicated with an asterisk. For the sake of simplicity, models with zero accuracy across all training stages are not included in the plot.*

The figure demonstrates that the leading model changes (i.e., the one with the highest score, marked with an asterisk) upon observing new pairs. In the first stage (top row), following the observation of the first pair (A > B), the top hypothesis is a Mallows model with the two symbols A and B, denoted as [AB]. All other hypotheses yield zero scores and are omitted. In the second stage, upon seeing the pair (B > C), two competing hypotheses emerge: a single Mallows model [ABC] versus the simultaneous presence of two Mallows models: [AB] and [BC]. However, the first hypothesis prevails as it boasts a higher score due to its reduced complexity. At the third stage, after observing the pair (D > E), the winning hypothesis is the concurrent existence of two Mallows models: [ABC] and [DE]. This reflects the point at which the observed symbols cannot be seamlessly integrated into a single ranking.

The following two stages illustrate the most crucial comparison in the study by (Nelli et al., 2023). In the fourth stage, after witnessing the pair (E > F), the most plausible hypothesis becomes the simultaneous presence of two Mallows models: [ABC] and [DEF]. This scenario mirrors the concurrent learning of two disconnected rankings, a phenomenon documented in (Nelli et al., 2023), before the introduction of the connecting pair (C > D). Finally, in the fifth stage, upon observing the connecting pair (C > D), the leading hypothesis transitions to a single Mallows model encompassing all the observed symbols, [ABCDE]. This result illustrates how the model selection approach can account for the rapid merger of two distinct rankings after the connecting pair is observed, as reported in (Nelli et al., 2023). Notably, this capability of the model reflects the ability of monkeys to correctly perform TI on previously disconnected pairs (e.g., AD) at the first trial after the introduction of the connecting pair (C > D), as observed in (Brunamonti et al., 2016).

It's important to note that, for the sake of clarity in this simulation, we've skipped the procedure outlined in the Methods section for learning the central point of each Mallows model. Consequently, the best model can be inferred after observing each pair just once. Inferring the optimal Mallows model and its central point simultaneously is entirely feasible and yields the same results but necessitates a larger amount of training data. This is consistent with the learning phase of TI experiments, where pairs are typically presented multiple times until a performance criterion is met.

### *Third simulation: neural-like codes supporting TI in the Mallows ranking model*

In the third simulation, we trained multilayer perceptrons (MLP) to develop neural-like codes that mimic Mallows models and can be compared with the neural data from (Nelli et al., 2023) and (Brunamonti et al., 2016). Note that our MLP implementation is not designed to reproduce specific anatomical features of neural circuits for TI (unlike previous studies, e.g., (Frank et al., 2003)) or to mimic the learning process of participants. Rather, we follow the approach of model-based neuroscience to use neural networks (not necessarily incorporating biologically realistic features) as "encoding models", to explore whether the distributed neural representations that emerge in such networks capture some aspects of the neural codes of biological organisms (Naselaris et al., 2011).

We compare two MLP networks trained on two different datasets, representing the critical comparison in (Nelli et al., 2023) between two learning stages. The first dataset corresponds to the observation of the pairs AB, BC, DE, EF (fourth training stage in Figure 5), while the second dataset is based on the subsequent observation of the connecting pair CD (fifth training stage in Figure 5). Following the results of the second simulation, the first dataset is sampled from a mixture of the two Mallows models [ABC, DEF], while the second dataset is sampled from a single Mallows model [ABCDEF], using the footrule distance as a metric.

We trained the two MLP networks until convergence. After training, we analyzed the internal representations of the symbols they developed in the space of their hidden layer activity after the fourth (Figure 6A,B) and fifth (Figure 6C,D) learning stages. To obtain the internal representation of each symbol, we provided the network with inputs where only the first, symbol-corresponding pattern (the first 6 elements of the input vector) was activated, while the second pattern of the input pair was inactivated (the last 6 elements of the input pattern were zeroed out).

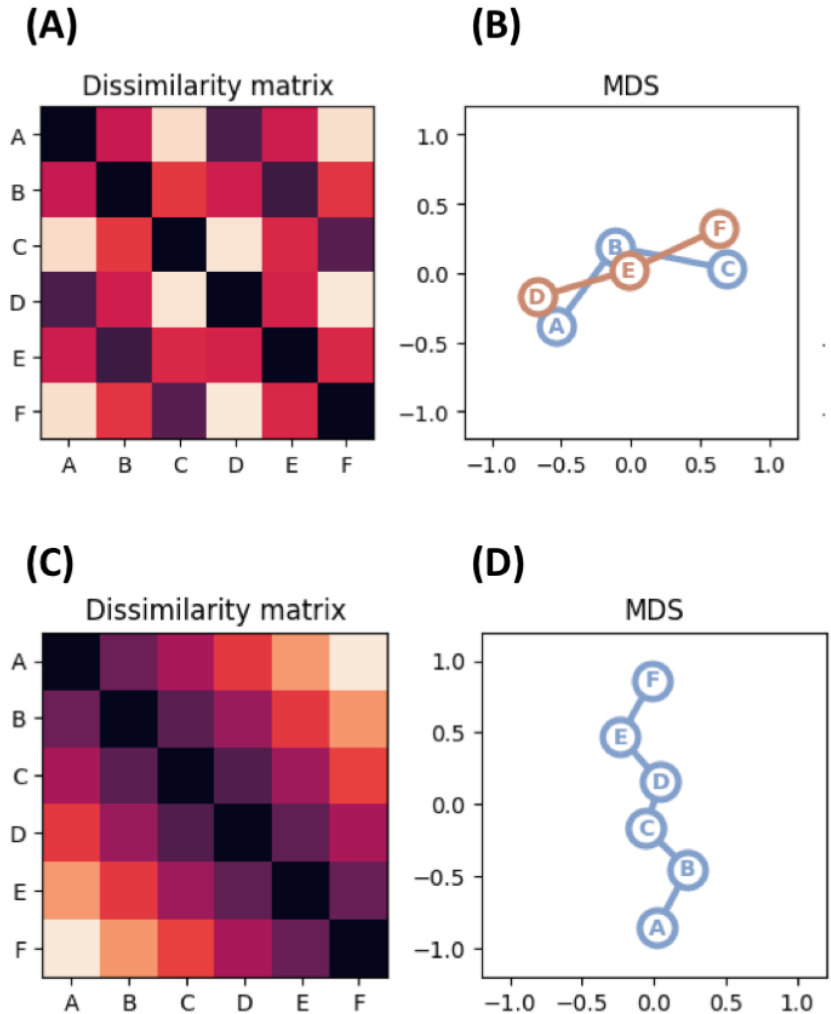

***Figure 6. Results of the third simulation: internal representations of symbols in the MLP networks trained to perform TI.*** *This figure shows the results of the comparison of neural-like representations developed by MLP networks trained in two learning stages: during the fourth learning stage (before the connecting pair CD is observed) and the fifth learning stage of the second simulation (after the connecting pair CD is observed). (A) Dissimilarities in internal representations of a neural network trained on pairs sampled from a mixture of the two Mallows models [ABC] and [DEF]. (B) 2-dimensional reduction of hidden representations of symbols after multidimensional scaling (MDS) of the matrix in panel A. (C) Dissimilarities in internal representations of a neural network trained on pairs sampled from the Mallows model [ABCDEF]. (D) 2-dimensional reduction of hidden representations of symbols after multidimensional scaling (MDS) of the matrix in panel C.*

Figure 6A,B illustrates that after the fourth learning stage, two rankings of three symbols each emerge in the hidden layer of the MLP network. Figure 6A displays the dissimilarity matrix between all symbols, defined as the Euclidean distance between the vectors of their corresponding hidden nodes. Figure 6B shows the projection of each symbol representation on a 2D space after dimensionality reduction by multidimensional scaling (MDS) of the matrix in Figure 6A. Symbols are arranged into two rankings, with the distances between representations in 2D space reflecting serial position. Additionally, symbols sharing the same ranking position between the two models (A and D, B and E, C and F) occupy nearly overlapping positions in the 2D space.

Figure 6C,D demonstrates that after the fifth learning stage, a single ranking of six symbols emerges in the hidden layer of the MLP network. Figure 6C displays the dissimilarity matrix between all symbols, while Figure 6D shows the projection of each symbol representation on a 2D space after dimensionality reduction by multidimensional scaling. The distances between representations in 2D space exemplify both the symbolic distance effect, as evidenced by symbols being organized along a line, and the serial position effect, as evidenced by the shorter distance between symbols C and D compared to the other distances. In summary, the results in Figure 6 indicate that MLPs trained to perform TI with Mallows models develop neural-like codes that closely resemble human frontoparietal networks (Nelli et al., 2023).

To further investigate whether the neural-like codes developed by the MLP networks are modulated by serial position and symbolic distance, we plotted their simulated neural activity as a function of serial position of input pairs (Figure 7A,C) and of symbolic distance (Figure 7B,D) in the two learning phases (Figure 7A,B for the fourth learning phase and Figure 7C,D for the fifth learning phase). After the fourth learning phase, the simulated activations of the MLP neural-like codes are too short to assess serial position effects (Figure 7A); however, they show a clear modulation by symbolic distance (Figure 7B). After the fifth learning phase, the simulated activations of the MLP neural-like codes are equivalent to the ranking probabilities of the Mallows model [ABCDEF], which supplied the dataset for this learning phase (see Supplementary Materials, Figure S3) and are modulated by both serial position (Figure 7C) and symbolic distance (Figure 7D). This latter result closely resembles the modulation of neural activity in monkey prefrontal areas by both serial position and symbolic distance (Brunamonti et al., 2016); compare with Figure 1.

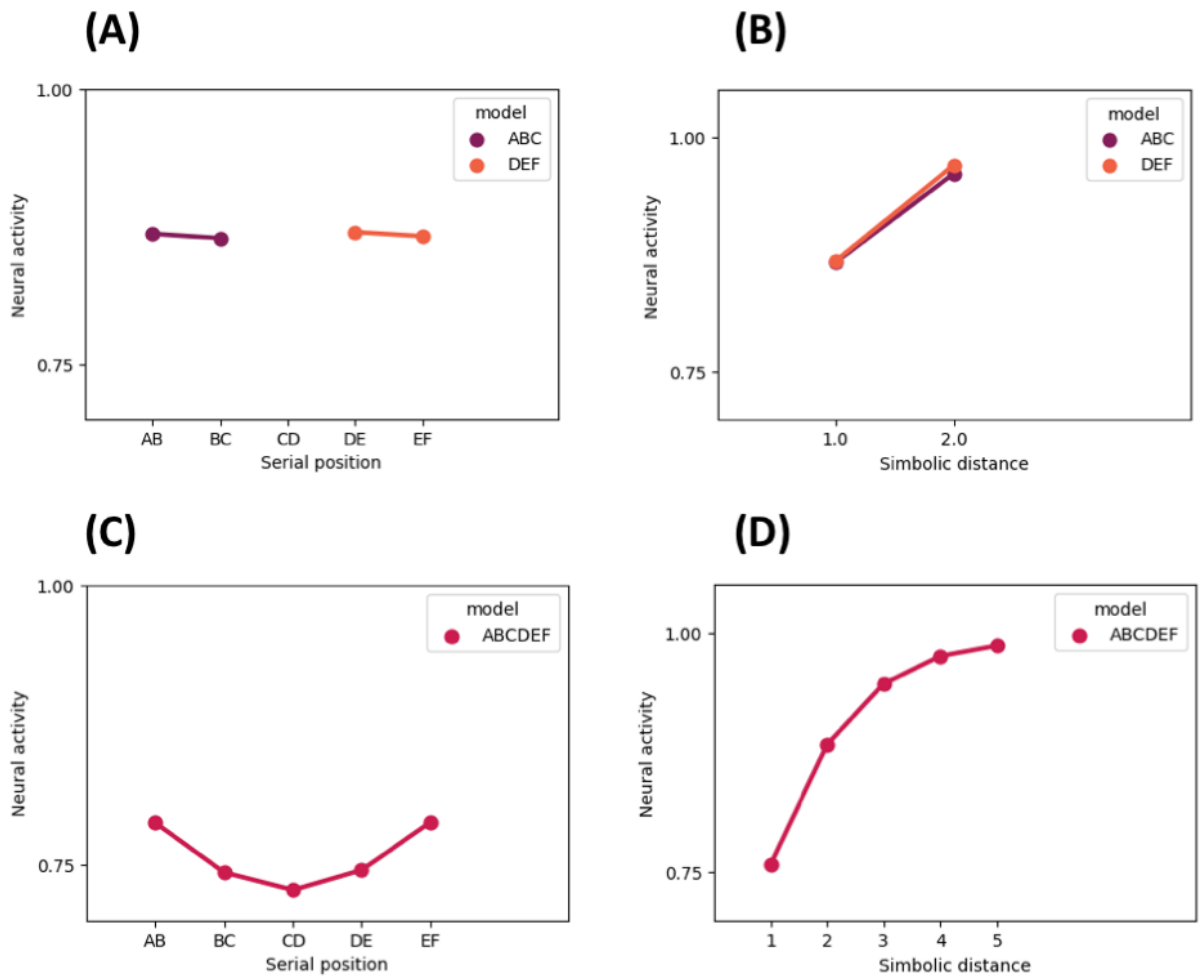

*Figure 7. Results of the third simulation: simulated activity in the MLP networks trained to perform TI. Simulated neural activity of the MLP trained after the first learning phase, based on A) the serial position of input pairs and B) the symbolic distance between symbols in input pairs. Simulated neural activity of the MLP trained after the second learning phase, based on C) the serial position of input pairs and D) the symbolic distance between symbols in input pairs.*

**Discussion**

Transitive Inference (TI) tasks serve as a valuable tool for evaluating the capacity of humans and various animal species to deduce novel relationships between pairs of items, even when those pairs have never been encountered together. This entails making inferences based on established relations among other pairs of items, exemplified by inferring that A > C, leveraging known relationships like A > B and B > C. Through extensive research, we have come to understand that several animals possess the cognitive aptitude to tackle TI tasks, with diverse neural mechanisms implicated in the process.

From a computational perspective, researchers have explored various ways to conceptualize TI. Some have approached it as a form of probabilistic inference that depends on a model or mental schema, one that systematically organizes the items into a cohesive relational structure, typically represented as a ranking (Jensen et al., 2013). An alternative approach has been to employ feedforward or recurrent neural networks trained to handle IT tasks, as evidenced by studies such as those by (Ciranka et al., 2022; Kay et al., 2024). Nevertheless, it is challenging to provide a concise, unified explanation for the multitude of behavioral and neural characteristics observed in TI. These encompass well-documented phenomena like the symbolic distance effect (SDE), the serial position effect (SPE), and the more recent discovery that the brain maintains distinct ranking models for items that have never been directly paired (e.g., the existence of separate ranks for AB and CD after observing A > B and C > D), yet has the ability to swiftly amalgamate them into a unified ranking model upon encountering a connecting pair (e.g., BC).

Here, we present a straightforward yet effective implementation of an inferential approach, using the Mallows ranking model. This model, built upon the concept of probabilistic inference, convincingly replicates the aforementioned phenomena. We present three simulations that not only substantiate the validity of this probabilistic ranking model but also underscore its elegance as an explanation for TI's behavioral outcomes and the neural underpinnings observed in both humans (Nelli et al., 2023) and monkeys (Brunamonti et al., 2016).

The first simulation serves as a compelling demonstration of the efficacy of our probabilistic ranking model. Remarkably, a model characterized by a sole parameter adeptly recapitulates TI's two fundamental effects: the symbolic distance effect (SDE) and the serial position effect (SPE). At a neurophysiological level, this model offers the intriguing proposition that the brain maintains probabilistic representations of rankings, subsequently harnessing them to gauge the likelihood of rank-order for both familiar and unfamiliar pairs (e.g., P(AB) and P(AF)).

In the second simulation, we take a significant step forward by expanding the traditional Mallows ranking model, introducing the concept of Bayesian model selection. This innovative approach effectively captures the brain's ability to formulate various hypotheses about rankings during the learning process. Furthermore, it elucidates how disparate rankings can be seamlessly consolidated into a single ranking when the brain encounters pairs featuring connecting symbols, as demonstrated in the research by (Nelli et al., 2023).

The third simulation emerges as a pivotal component of our study, offering insight into the internal neural-like codes that a neural network acquires when trained to mimic the Mallows model. Significantly, these codes align with neural recordings taken during TI. They offer a remarkable account of both the SDE and SPE, showcasing their influence on prefrontal population neural activity in monkeys, as observed in the work by (Brunamonti et al., 2016). Moreover, these neural-like codes effectively encapsulate the changes in prefrontal neural populations that occur during the process of connecting previously disjointed rankings, as documented in the study by (Nelli et al., 2023). This third simulation, while speculative, strongly suggests that the probabilistic approach inherent to the Mallows model might transcend mere mathematical convenience, potentially furnishing deeper insights into the neural computations underpinning TI. However, it is essential to acknowledge the need for further studies to validate this speculative proposal.

Looking at this model from a computational standpoint unveils a distinct advantage over alternative proposals. It possesses the unique capability to concurrently reproduce several key effects that were hitherto challenging to reconcile within a single model. What sets this model apart is its inherent parsimony, reliant on a solitary parameter - precision ($\beta$). This stands in stark contrast to the Betasort model, which necessitates the specification of separate parameters for encoding beta distributions associated with each symbol, as outlined by (Jensen et al., 2015). Notably, in comparison to prior neural network models, such as the one proposed by (Kay et al., 2024), our model not only provides a more comprehensive account of various aspects of TI tasks but also offers a heightened level of functional interpretability, rooted in probabilistic computations. Our model assumes that the same ranking-based, inferential mechanisms are used to process both seen pairs (e.g., AB) and unseen pairs (e.g., AE). If the brain used two distinct mechanisms for seen and unseen pairs – say, associative mechanisms for known pairs and inferential mechanisms only for novel pairs – then one would expect seen pairs to be processed faster and more accurately than unseen pairs. On the contrary, as discussed in the introduction, the majority of TI studies report that unseen pairs with greater symbolic distance are processed faster and more accurately than seen pairs with smaller symbolic distance. While associative mechanisms might coexist with inferential mechanisms, it is unlikely that the former are sufficient to explain the results of TI studies.

Another element of novelty of our study compared to previous inferential models of TI is the use of Bayesian model selection to score different hypotheses about rankings during learning. This method provides a novel explanation of how participants might learn (and merge) rankings, whose number and length is unknown a priori. In our simulations, we have assumed that the hypothesis that best balances accuracy and complexity (as measured by BIC) is the one that is selected in practice and this permits to account for the finding that the brain rapidly connects rankings upon encountering connecting pairs (Nelli et al., 2023). However, various studies reported individual differences in a propensity to integrate the pairs and therefore ability to perform TI in novel pairs (e.g., humans: (Lazareva and Wasserman, 2010); pigeons: (Lazareva and Wasserman, 2012)). Other studies reported species differences in readiness to integrate (e.g., (Maclean et al., 2008), comparing lemur species, and (Bond et al., 2010), comparing corvid species). From the perspective of our model, these results would indicate individual or species differences in the ways the hypotheses are scored or selected, meaning that in some cases suboptimal hypotheses (from the perspective of BIC) might be adopted. For instance, by considering that selecting complex hypotheses entails some cognitive cost, some species might privilege simpler hypotheses, even if they entail some loss of accuracy. Whether this perspective might help advance our understanding of individual and species differences during TI remains to be assessed by future studies.

From a neural perspective, our proposition that the brain employs a probabilistic ranking model, serving as a relational schema, aligns with the Bayesian brain hypothesis. This hypothesis posits that cognitive processing can be fundamentally understood as a form of Bayesian inference over a generative model, capturing the statistical regularities inherent in cognitive tasks. The brain, as a probabilistic inference machine, could potentially use various mechanisms for performing this task. These include the utilization of probabilistic population codes (Pouget et al., 2013), the implementation of sampling methods (Fiser et al., 2010) or other approximate (e.g., variational (Friston, 2010; Parr et al., 2022)) schemes. Our Figure 4 emphasizes that, within the scope of our task, deploying the complete Mallows distribution or resorting to a sampling approach yields notably similar results. Consequently, while our simulations do not contribute to the ongoing debate about the exact neural mechanisms at play, they underscore the robustness of the model in capturing essential aspects of TI. In sum, while previous research suggests that the brain might encode probabilistic variables, it remains to be assessed whether and how exactly it forms probability distributions over rankings.

A fundamental premise of our approach centers on the idea that the human brain integrates experiences within a relational schema that affords inference. When it comes to transitive inference, the task at hand only necessitates relatively simple, one-dimensional relational structures in the form of rankings, facilitating the learning and inference of preference distributions. Other cognitive tasks, such as spatial navigation, may demand more complex and higher-dimensional structures akin to cognitive maps (Behrens et al., 2018; Bellmund et al., 2018; George et al., 2021; Stoianov et al., 2022). Within the scope of our model, we assume that individuals, including humans and other animals, performing transitive inference have already acquired through their lifetimes an a priori knowledge of the correct generative model's form – in this case, a ranking model like the Mallows model. Our model is designed to account for an individual's cognitive processes at a relatively high level of analysis; namely, the "computational" level in the famous distinction between the computational, the algorithmic, and the implementational levels introduced by (Marr, 1982). However, various brain structures, such as the hippocampal formation and the prefrontal cortex, have the capacity to encode rich internal models of sequences, which might support the inferential mechanisms proposed by our model (Bellet et al., 2024; Buzsáki and Tingley, 2018; Xie et al., 2022). Mapping the computational model proposed here to a specific neural substrate is beyond the scope of this study and remains an open objective for future research. Furthermore, future studies could consider extensions of our computational model, which could involve the capability to select the most likely form of the generative model following observations of the data distribution, as exemplified in the work by (Kemp and Tenenbaum, 2008).

Notably, the Bayesian modeling approach has proven to be a valuable tool in investigating the neural substrates implicated in transitive inference. This approach has been successfully applied in studies focusing on critical brain structures, including the hippocampus (Dragoi and Buzsáki, 2006; Lisman and Redish, 2009; Liu et al., 2018; Penny et al., 2013; Pezzulo et al., 2014; Stoianov et al., 2022; Ujfalussy and Orbán, 2022), as well as the prefrontal cortex (Barceló, 2021; Donnarumma et al., 2016; Stoianov et al., 2015). Our work demonstrates that it is possible to emulate the probabilistic representations inherent in the Mallows model through the utilization of a simple feedforward neural network. However, it is important to emphasize that our choice of this neural network is primarily motivated by the ease of training and should not be misconstrued as a representation of biological plausibility. The ways the brain learns and encodes probabilistic representations remain a topic of fervent debate within the scientific community.

In our study, we specifically employed the Spearman's footrule as the metric for measuring the distance between rankings. What is intriguing is that our use of both the Spearman's footrule and the Spearman's rank correlation allowed us to successfully replicate the symbolic distance effect (SDE) and the serial position effect (SPE). In contrast, employing other widely-used metrics, such as Kendall's tau, Hamming, or Ulam distances, only reproduced the SDE without accounting for the SPE effect, as outlined in the Supplementary Materials. The unique strength of the Spearman's footrule and the Spearman's rank correlation lies in their ability to incorporate information regarding the distances within ranks between symbols. This feature aligns with the notion that the brain retains a representation of this distance when making pairwise comparisons (Brunamonti et al., 2016). Other distances, such as the Kendall's tau, in which all rankings with arbitrarily swapped pairs have the same distance from the central point (possibly simulating failures of memory for the ordering of symbols) can reproduce the symbolic distance effect but not the serial position effect, see the Supplementary Materials. This result suggests that memory issues, such as forgetting the ordering of learning pairs, is unlikely to fully explain the results of TI tasks. It is noteworthy that there exists the potential to introduce non-linearity into Spearman's distances between symbols, in which greater (or smaller) distances hold varying levels of influence in distinguishing between two permutations. This can be accomplished by simply adding an additional parameter to the metrics, as detailed in the Supplementary Materials. Subsequent research endeavors may aim to explore whether these non-linearities can enhance the quantitative fit with experimental data related to transitive inference (TI).

By casting transitive inference as probabilistic preference learning, our model makes novel predictions that could be tested in future studies. First, in our model, processing pairs, both seen and unseen, requires using an entire ranking, not just the pairs under examination. This is because the probability of a pair results from an inference (or a sampling process) over the entire Mallows model. Future studies might address whether at the neural level, transitive inference elicits the entire ranking sequence versus the single pair under examination. Second, our model assumes that the neural representation supporting transitive inference is probabilistic and that transitive inference considers a probability distribution over rankings, rather than over the values of single stimuli in the ranking, as assumed for example by the Betasort model (Jensen et al., 2015). In principle, future studies might assess whether transitive inference elicits probabilistic representations of the kind required by the Mallows model and in that case, the specific scheme (e.g., sampling, variational approximation) it uses. Third, assuming that the beta parameter shown in Figure 2 (measuring the precision of the probabilistic representation) could be learned over time, its value would become very small after extensive training (or overtraining) – then the probability distribution would collapse to its central point, with very little variance. Our model predicts that if this happens, both the symbolic distance and (especially) the serial position effects would be largely attenuated, as the probability of all pairs will reach ceiling levels. Future studies employing extensive training regimes might test whether these predicted changes at the behavioral and the neural levels hold. Fourth, our model suggests at least two possible reasons for failures of transitive inference. The former reason for failure of transitive inference is a high value of the beta parameter shown in Figure 2, possibly reflecting learning errors or cognitive deficits, which could lead to high variance in the Mallows model or even the failure to correctly learn the true central point of the distribution. Studies that address the putative neural representations of the beta parameter might be able to test whether high levels of beta are associated with failures of transitive inference. The latter reason for failures of transitive inference is employing a sampling scheme with few samples (Fiser et al., 2010; Vul et al., 2014). Assuming that sampling has an associated cognitive cost, studies that manipulate cognitive load during transitive inference could assess whether the reduced availability of cognitive resources translates in a decreased performance and whether in turn this decrease is well explained by a Mallows model that employs few samples.

One limitation of the current model is that it addresses choices, but not other response measures such as reaction times. However, it is possible to derive response times from the probabilities that the model assigns to pairs, by considering that reaction times scale as a (linear) function of the absolute value of (a transformation of) response probabilities (Bonnet et al., 2008). Considering this (linear) relation between response probabilities and response times would allow reproducing a common finding across TI studies, that pairs with greater (smaller) symbolic distance are processed not only more accurately but also faster. However, providing a quantitative validation of model-derived reaction times is an open objective for future research. Another intriguing avenue for future research involves assessing the extent to which our approach can effectively model the trial-by-trial learning dynamics observed during TI studies. In this present work, we chose to adopt a simplified learning protocol, where each pair was presented only once, and the centers of the Mallows distributions were assumed to be known. However, our approach inherently lends itself to the simultaneous learning of the centers of these distributions and the process of model selection. Future research could delve deeper into the investigation of how well this method accounts for the intricacies of trial-by-trial pairwise comparisons, a prominent feature of TI studies.

## Acknowledgements

This research received funding from the European Union's Horizon 2020 Framework Programme for Research and Innovation under the Specific Grant Agreement No. 952215 (TAILOR); the European Research Council under the Grant Agreement No. 820213 (ThinkAhead), the Italian National Recovery and Resilience Plan (NRRP), M4C2, funded by the European Union – NextGenerationEU (Project IR0000011, CUP B51E22000150006, "EBRAINS-Italy"; Project PE0000013, "FAIR"; Project PE0000006, "MNESYS"), and the PRIN PNRR P20224FESY. The GEFORCE Quadro RTX6000 and Titan GPU cards used for this research were donated by the NVIDIA Corporation. The funders had no role in study design, data collection and analysis, decision to publish, or preparation of the manuscript.

## Data statement

The data and materials for all experiments are available at https://github.com/francesco-mannella/Transitive-Inference-As-Probabilistic-Preference-Learning

# References

Barceló, F., 2021. A Predictive Processing Account of Card Sorting: Fast Proactive and Reactive Frontoparietal Cortical Dynamics during Inference and Learning of Perceptual Categories. J Cogn Neurosci 33, 1636–1656. https://doi.org/10.1162/jocn_a_01662

Behrens, T.E.J., Muller, T.H., Whittington, J.C.R., Mark, S., Baram, A.B., Stachenfeld, K.L., Kurth-Nelson, Z., 2018. What Is a Cognitive Map? Organizing Knowledge for Flexible Behavior. Neuron 100, 490–509. https://doi.org/10.1016/j.neuron.2018.10.002

Bellet, M.E., Gay, M., Bellet, J., Jarraya, B., Dehaene, S., van Kerkoerle, T., Panagiotaropoulos, T.I., 2024. Spontaneously emerging internal models of visual sequences combine abstract and event-specific information in the prefrontal cortex. Cell Reports 43, 113952. https://doi.org/10.1016/j.celrep.2024.113952

Bellmund, J.L.S., Gärdenfors, P., Moser, E.I., Doeller, C.F., 2018. Navigating cognition: Spatial codes for human thinking. Science 362, eaat6766. https://doi.org/10.1126/science.aat6766

Bond, A.B., Kamil, A.C., Balda, R.P., 2003. Social complexity and transitive inference in corvids. Animal Behaviour 65, 479–487. https://doi.org/10.1006/anbe.2003.2101

Bond, A.B., Wei, C.A., Kamil, A.C., 2010. Cognitive representation in transitive inference: a comparison of four corvid species. Behav Processes 85, 283–292. https://doi.org/10.1016/j.beproc.2010.08.003

Bonnet, C., Fauquet Ars, J., Estaún Ferrer, S., 2008. Reaction times as a measure of uncertainty. Psicothema 20, 43–48.

Botvinick, M., Watanabe, T., 2007. From Numerosity to Ordinal Rank: A Gain-Field Model of Serial Order Representation in Cortical Working Memory. J. Neurosci. 27, 8636–8642. https://doi.org/10.1523/JNEUROSCI.2110-07.2007

Brunamonti, E., Mione, V., Di Bello, F., Pani, P., Genovesio, A., Ferraina, S., 2016. Neuronal Modulation in the Prefrontal Cortex in a Transitive Inference Task: Evidence of Neuronal Correlates of Mental Schema Management. J Neurosci 36, 1223–1236. https://doi.org/10.1523/JNEUROSCI.1473-15.2016

Bryant, P.E., Trabasso, T., 1971. Transitive Inferences and Memory in Young Children. Nature 232, 456–458. https://doi.org/10.1038/232456a0

Burt, C., 1909. Experimental Tests of General Intelligence. British Journal of Psychology 3, 94–177.

Buzsáki, G., Tingley, D., 2018. Space and Time: The Hippocampus as a Sequence Generator. Trends Cogn. Sci. (Regul. Ed.) 22, 853–869. https://doi.org/10.1016/j.tics.2018.07.006

Ciranka, S., Linde-Domingo, J., Padezhki, I., Wicharz, C., Wu, C.M., Spitzer, B., 2022. Asymmetric reinforcement learning facilitates human inference of transitive relations. Nat Hum Behav 6, 555–564. https://doi.org/10.1038/s41562-021-01263-w

Davis, H., 1992. Transitive inference in rats (Rattus norvegicus). J Comp Psychol 106, 342–349. https://doi.org/10.1037/0735-7036.106.4.342

De Lillo, C., Floreano, D., Antinucci, F., 2001. Transitive choices by a simple, fully connected, backpropagation neural network: implications for the comparative study of transitive inference. Anim.Cogn. 4, 61–68. https://doi.org/10.1007/s100710100092

Di Antonio, G., Raglio, S., Mattia, M., 2023. Ranking and serial thinking: A geometric solution. https://doi.org/10.1101/2023.08.03.551859

Diaconis, P., 1988. Group representations in probability and statistics. Lecture notes-monograph series 11, i–192.

Donnarumma, F., Maisto, D., Pezzulo, G., 2016. Problem Solving as Probabilistic Inference with Subgoaling: Explaining Human Successes and Pitfalls in the Tower of Hanoi. PLOS Computational Biology 12, e1004864. https://doi.org/10.1371/journal.pcbi.1004864

Dragoi, G., Buzsáki, G., 2006. Temporal encoding of place sequences by hippocampal cell assemblies. Neuron 50, 145–157. https://doi.org/10.1016/j.neuron.2006.02.023

Dusek, J.A., Eichenbaum, H., 1997. The hippocampus and memory for orderly stimulus relations. Proc Natl Acad Sci U S A 94, 7109–7114. https://doi.org/10.1073/pnas.94.13.7109

Fiser, J., Berkes, P., Orbán, G., Lengyel, M., 2010. Statistically optimal perception and learning: from behavior to neural representations. Trends Cogn Sci 14, 119–130. https://doi.org/10.1016/j.tics.2010.01.003

Frank, M.J., Rudy, J.W., O'Reilly, R.C., 2003. Transitivity, flexibility, conjunctive representations, and the hippocampus. II. A computational analysis. Hippocampus 13, 341–354.

Friston, K.J., 2010. The free-energy principle: a unified brain theory? Nat Rev Neurosci 11, 127–138. https://doi.org/10.1038/nrn2787

George, D., Rikhye, R.V., Gothoskar, N., Guntupalli, J.S., Dedieu, A., Lázaro-Gredilla, M., 2021. Clone-structured graph representations enable flexible learning and vicarious evaluation of cognitive maps. Nat Commun 12, 2392. https://doi.org/10.1038/s41467-021-22559-5

Jensen, G., Alkan, Y., Ferrera, V.P., Terrace, H.S., 2019. Reward associations do not explain transitive inference performance in monkeys. Science Advances 5, eaaw2089. https://doi.org/10.1126/sciadv.aaw2089

Jensen, G., Altschul, D., Danly, E., Terrace, H., 2013. Transfer of a Serial Representation between Two Distinct Tasks by Rhesus Macaques. PLOS ONE 8, e70285. https://doi.org/10.1371/journal.pone.0070285

Jensen, G., Muñoz, F., Alkan, Y., Ferrera, V.P., Terrace, H.S., 2015. Implicit Value Updating Explains Transitive Inference Performance: The Betasort Model. PLOS Computational Biology 11, e1004523. https://doi.org/10.1371/journal.pcbi.1004523

Kay, K., Biderman, N., Khajeh, R., Beiran, M., Cueva, C.J., Shohamy, D., Jensen, G., Wei, X.-X., Ferrera, V.P., Abbott, L.F., 2024. Emergent neural dynamics and geometry for generalization in a transitive inference task. PLoS Comput Biol 20, e1011954. https://doi.org/10.1371/journal.pcbi.1011954

Kemp, C., Tenenbaum, J.B., 2008. The discovery of structural form. Proceedings of the National Academy of Sciences 105, 10687–10692.

Lazareva, O.F., Kandray, K., Acerbo, M.J., 2015. Hippocampal lesion and transitive inference: dissociation of inference-based and reinforcement-based strategies in pigeons. Hippocampus 25, 219–226. https://doi.org/10.1002/hipo.22366

Lazareva, O.F., Wasserman, E.A., 2012. Transitive inference in pigeons: Measuring the associative values of Stimuli B and D. Behavioural Processes 89, 244–255. https://doi.org/10.1016/j.beproc.2011.12.001

Lazareva, O.F., Wasserman, E.A., 2010. Nonverbal transitive inference: Effects of task and awareness on human performance. Behav Processes 83, 99–112. https://doi.org/10.1016/j.beproc.2009.11.002

Lisman, J., Redish, A.D., 2009. Prediction, Sequences and the Hippocampus. Phil. Trans. R. Soc. B 364, 1193–1201. https://doi.org/10.1098/rstb.2008.0316

Liu, K., Sibille, J., Dragoi, G., 2018. Generative Predictive Codes by Multiplexed Hippocampal Neuronal Tuplets. Neuron 99, 1329-1341.e6. https://doi.org/10.1016/j.neuron.2018.07.047

Maclean, E.L., Merritt, D.J., Brannon, E.M., 2008. Social Complexity Predicts Transitive Reasoning in Prosimian Primates. Anim Behav 76, 479–486. https://doi.org/10.1016/j.anbehav.2008.01.025

Mallows, C.L., 1957. Non-null ranking models. I. Biometrika 44, 114–130.

Marr, D., 1982. Vision: A Computational Investigation into the Human Representation and Processing of Visual Information. Henry Holt and Co., Inc., New York, NY, USA.

McGonigle, B.O., Chalmers, M., 1977. Are monkeys logical? Nature 267, 694–696. https://doi.org/10.1038/267694a0
24

Merritt, D.J., Terrace, H.S., 2011. Mechanisms of inferential order judgments in humans (Homo sapiens) and rhesus monkeys (Macaca mulatta). J Comp Psychol 125, 227–238. https://doi.org/10.1037/a0021572

Naselaris, T., Kay, K.N., Nishimoto, S., Gallant, J.L., 2011. Encoding and decoding in fMRI. Neuroimage 56, 400–410. https://doi.org/10.1016/j.neuroimage.2010.07.073

Nelli, S., Braun, L., Dumbalska, T., Saxe, A., Summerfield, C., 2023. Neural knowledge assembly in humans and neural networks. Neuron 111, 1504-1516.e9. https://doi.org/10.1016/j.neuron.2023.02.014

Parr, T., Pezzulo, G., Friston, K.J., 2022. Active Inference: The Free Energy Principle in Mind, Brain, and Behavior. MIT Press, Cambridge, MA, USA.

Penny, W.D., Zeidman, P., Burgess, N., 2013. Forward and Backward Inference in Spatial Cognition. PLoS Comput Biol 9, e1003383. https://doi.org/10.1371/journal.pcbi.1003383

Pezzulo, G., van der Meer, M.A.A., Lansink, C.S., Pennartz, C.M.A., 2014. Internally generated sequences in learning and executing goal-directed behavior. Trends in Cognitive Sciences 18, 647–657. https://doi.org/10.1016/j.tics.2014.06.011

Piaget, J., 1947. La psychologie de l'intelligence. [The psychology of intelligence.], La psychologie de l'intelligence. Armand Colin, Oxford, England. https://doi.org/10.4324/9780203278895

Pouget, A., Beck, J.M., Ma, W.J., Latham, P.E., 2013. Probabilistic brains: knowns and unknowns. Nature neuroscience 16, 1170–1178.

Schwarz, G., 1978. Estimating the dimension of a model. The annals of statistics 461–464.

Siemann, M., Delius, J.D., 1998. Algebraic learning and neural network models for transitive and non-transitive responding. European Journal of Cognitive Psychology 10, 307–334. https://doi.org/10.1080/713752279

Stoianov, I., Genovesio, A., Pezzulo, G., 2015. Prefrontal Goal Codes Emerge as Latent States in Probabilistic Value Learning. Journal of cognitive neuroscience 28, 140–157.

Stoianov, I., Maisto, D., Pezzulo, G., 2022. The hippocampal formation as a hierarchical generative model supporting generative replay and continual learning. Prog Neurobiol 217, 102329. https://doi.org/10.1016/j.pneurobio.2022.102329

Treichler, F.R., Raghanti, M.A., Van Tilburg, D.N., 2003. Linking of serially ordered lists by macaque monkeys (Macaca mulatta): list position influences. J Exp Psychol Anim Behav Process 29, 211–221. https://doi.org/10.1037/0097-7403.29.3.211

Treichler, F.R., Van Tilburg, D., 1996. Concurrent conditional discrimination tests of transitive inference by macaque monkeys: list linking. J Exp Psychol Anim Behav Process 22, 105–117.

Ujfalussy, B.B., Orbán, G., 2022. Sampling motion trajectories during hippocampal theta sequences. eLife 11, e74058. https://doi.org/10.7554/eLife.74058

Vasconcelos, M., 2008. Transitive inference in non-human animals: An empirical and theoretical analysis. Behavioural Processes 78, 313–334. https://doi.org/10.1016/j.beproc.2008.02.017

Vul, E., Goodman, N., Griffiths, T.L., Tenenbaum, J.B., 2014. One and done? Optimal decisions from very few samples. Cognitive science 38, 599–637.

Wei, C.A., Kamil, A.C., Bond, A.B., 2014. Direct and relational representation during transitive list linking in pinyon jays (Gymnorhinus cyanocephalus). J Comp Psychol 128, 1–10. https://doi.org/10.1037/a0034627

Xie, Y., Hu, P., Li, J., Chen, J., Song, W., Wang, X.-J., Yang, T., Dehaene, S., Tang, S., Min, B., Wang, L., 2022. Geometry of sequence working memory in macaque prefrontal cortex. Science 375, 632–639. https://doi.org/10.1126/science.abm0204

Xu, B., Wu, J., Xiao, H., Münte, T.F., Ye, Z., 2024. Inferior parietal cortex represents relational structures for explicit transitive inference. Cereb Cortex 34, bhae137. https://doi.org/10.1093/cercor/bhae137

Zeithamova, D., Dominick, A.L., Preston, A.R., 2012. Hippocampal and ventral medial prefrontal activation during retrieval-mediated learning supports novel inference. Neuron 75, 168–179. https://doi.org/10.1016/j.neuron.2012.05.010

# Supplementary materials

# Transitive inference as probabilistic preference learning

Francesco Mannella, Giovanni Pezzulo

*Institute of Cognitive Sciences and Technologies, National Research Council, Rome, Italy*

**Supplementary Tables**

The simulations discussed in the main text are based on the Spearman's footrule distance. Here we discuss alternative distances that are widely used for Mallows probability distributions: the Kendall's tau distance, the Ulam distance, the Hamming distance the Cayley distance, see Table S1.

| | |
|---|---|
| **Kendall's tau** | Minimum number of pairwise adjacent transpositions taking $\tau$ to $\mathbf{c}$. |
| **Ulam** | $N$ - length of longest increasing subsequence in $\tau\mathbf{c}^{-1} = \{(c_1, \tau_1), (c_2, \tau_2), \ldots, (c_N, \tau_N)\}$ |
| **Hamming** | $\#\{i : \tau_i \neq c_i\}$ |
| **Cayley** | minimum number of transpositions taking $\tau$ to $\mathbf{c}$. |
| **Spearman's footrule** | $\sum_i |\tau_i - c_i|$ |
| **Spearman's rank correlation** | $\sum_i (\tau_i - c_i)^2$ |

**Table S1**: Some of the most used metrics for Mallows probability distributions of ranking sequences; modified from (Diaconis, 1988).

**Supplementary Figures**

The selection of a specific metric for constructing the probability distribution exerts a significant influence on the pairwise decisions concerning relative rankings. This is because different metrics define the relationships between tokens in a sequence. Notably, metrics like Kendall's tau, Ulam, Hamming, or Cayley, which take into account more qualitative characteristics when comparing sequences, result in the emergence of the symbolic distance effect (SDE), but they do not adequately capture the serial position effect (SPE) (as illustrated in Figure S1A). In contrast, metrics such as Spearman's footrule, which are founded on quantitative distinctions in rank positions between sequences, yield decisions that are influenced by both the SDE and the SPE effects (as depicted in Figure S1B).

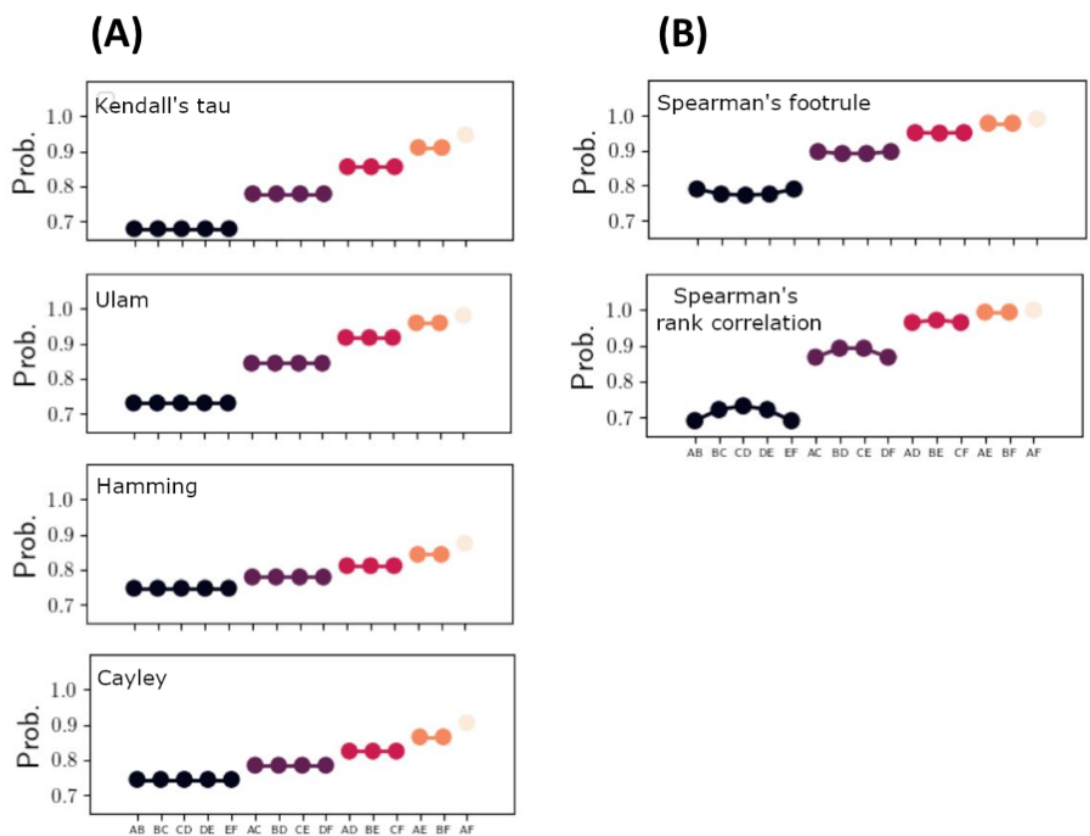

*Figure S1: Comparative Simulations of Relative-Rank Decisions using Mallows Models based on Different Metrics. Each plot presented in this figure depicts the probabilities associated with relative-rank decisions for every pair of tokens within a sequence of six ranks, considering models based on various metrics. (A) Mallows models grounded on metrics like Kendall's tau, Ulam, Hamming distance, and Cayley. Notably, these models do not successfully capture the Serial Position Effect (SPE). (B) Mallows models relying on Spearman's footrule and Spearman's rank correlation metrics. It's worth noting that these models effectively reproduce distinct manifestations of the SPE, highlighting the significance of metric choice in modeling the Serial Position Effect.*

A parameterized extension of Spearman's metrics, referred to as the Spearman Parametric Distance (SPD), can be derived.

$$SPD = \sum_i |\tau_i - c_i|^\gamma$$

With this derived metric, we can explore the $\gamma$ parameter space to identify a metric that aligns more effectively with the experimental data pertaining to the SPE effect (refer to Figure S2).

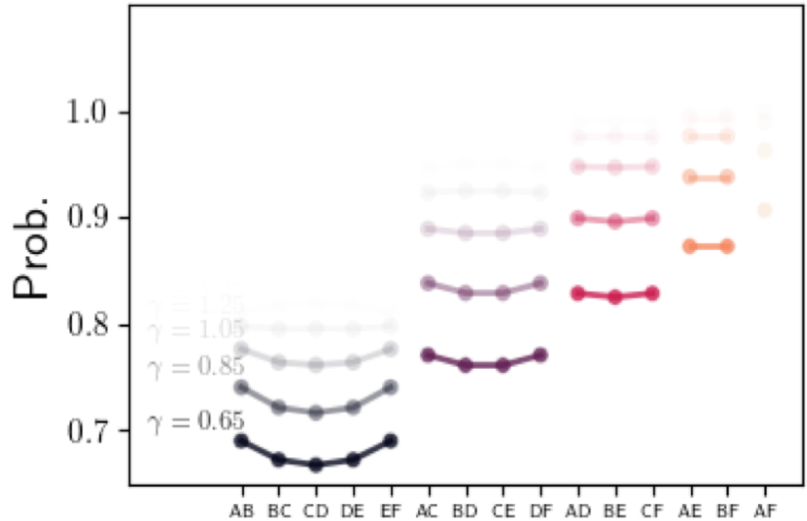

*Figure S2: Simulations of relative-rank decisions using a Mallows model based on the Spearman Parametric Distance. Various simulations employing different values of the parameter $\gamma$ are presented. The optimal value is identified as the least translucent/most distinct data series.*

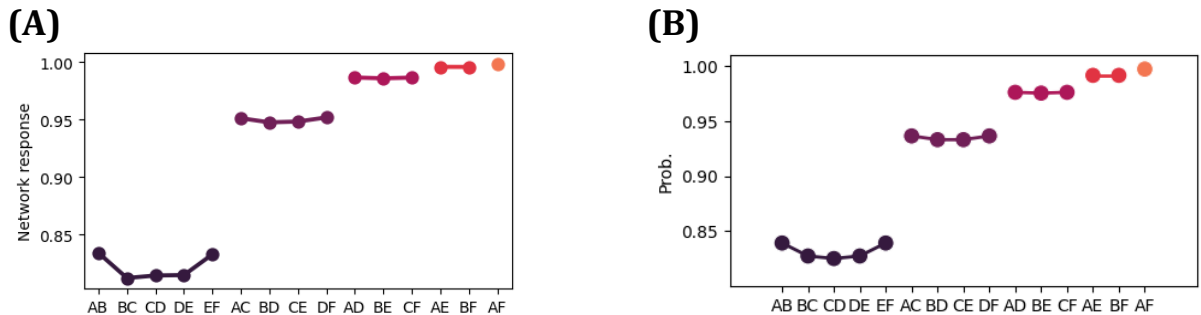

*Figure S3: Simulated activity in an MLP network trained for TI is compared to the probability distribution of relative-rank decisions using a Mallows model. (A) The MLP network's responses to various symbol pairs after training with a Mallows model guided by Spearman's footrule. (B) Probabilities of ranking responses based on the original Mallows model.*